\documentclass[aps,showpacs,pra,twocolumn,groupedaddress]{revtex4}
\usepackage{mathptmx}
\usepackage{graphicx}
\usepackage{dcolumn}
\usepackage{bm}
\usepackage{color}

\begin{document}

\title{Coulomb focusing at above-threshold ionization in elliptically\\ polarized mid-infrared strong laser fields}

\author{Chengpu Liu}
\email{Chengpu.Liu@mpi-hd.mpg.de}
\affiliation{Max-Planck-Institut f\"ur Kernphysik, Saupfercheckweg 1,  Heidelberg 69117, Germany}

\author{Karen Z. Hatsagortsyan}
\email{K.Hatsagortsyan@mpi-k.de}
\affiliation{Max-Planck-Institut f\"ur Kernphysik, Saupfercheckweg 1,   Heidelberg 69117, Germany}

\date{\today}

\begin{abstract}

The role of Coulomb focusing in above-threshold ionization in an elliptically polarized mid-infrared strong laser field is investigated within a semiclassical model  incorporating tunneling and Coulomb field effects. It is shown that Coulomb focusing up to moderate ellipticity values ($\xi\lesssim 0.3$) is dominated by multiple forward scattering of the ionized electron by the atomic core that creates a characteristic low-energy structure in the photoelectron spectrum and is responsible for the peculiar energy scaling of the ionization normalized yield along the major polarization axis. At higher ellipticities, the electron continuum dynamics is disturbed by the Coulomb field effect mostly at the  exit of the ionization tunnel. 
Due to the latter, the normalized yield is found to be enhanced, with the enhancement factor being sharply pronounced at intermediate ellipticities.

\end{abstract}

\maketitle

\section{Introduction}\label{intro}

The rescattering of an ionized electron by the atomic core in a strong laser field plays a fundamental role in strong field physics \cite{Schafer_1993,CorkumPRL93}, in particular, giving rise to above-threshold ionization, high-order harmonic generation and nonsequential multiple ionization \cite{Becker_review,Agostini_review,Walker,Moshammer,Feuerstein,DiChiara}. In a laser field of elliptical polarization, the ionized electron acquires a lateral drift motion with respect to the major polarization axis (m.p.a.) which could seem to suppress the rescattering, with the consequence of extinguishing the related effects. However, recently it has been shown that, even in a laser field of elliptical polarization, the rescattering can play a significant role in nonsequential double-ionization \cite{Shvetsov,Eberly,Wang,Uzer}. Nonsequential double-ionization is due to back-scattering of the  electron at small impact parameters. In contrast, the multiple forward-scattering of an ionized electron by the atomic core at large impact parameters induces Coulomb focusing (CF)   \cite{BrabecIvanov,YudinIvanov,Villeneuve}. Another contribution to CF comes from the initial Coulomb disturbance [the momentum transfer to the ionizing electron by a Coulomb field
at the initial part of the electron trajectory near the tunnel exit] \cite{Shvetsov-Shilovski}. Recent experiments \cite{DiMauroNP09,Catoire,XuPRL09} have shown that a characteristic spike-like low-energy structure (LES) arises in the energy distribution of electrons emitted along the polarization direction of linearly polarized mid-infrared laser radiation. The multiple forward scattering is shown to be responsible for LES  \cite{Liu,Yan,Liu_JPB} (for other aspects see \cite{Telnov,Chu,Rost,Burgdorfer}). A question arises if the  multiple forward scattering of the rescattering electron survives in a field of elliptical polarization and, in general, how CF and LES are modified due to ellipticity of the field.

Several experiments have been devoted to above-threshold ionization in an elliptically polarized laser field \cite{Bucksbaum,Bashkansky,Paulus_98,Paulus_00}, in particular, the dodging phenomenon has been discovered \cite{Paulus_98,Paulus_00,Becker_Paulus}. It is expressed as a quick drop of the photoelectron normalized yield  along m.p.a. with increasing ellipticity [the yield is normalized to the one in a linearly polarized field] that is reversed when circular polarization is approached. The first indication of the Coulomb field effects for above-threshold ionization in an elliptically polarized 
field, manifested in the lack of the four-fold symmetry of the photoelectron angular distribution, has been shown  experimentally \cite{Bashkansky} and discussed theoretically in \cite{Mittleman,Ehlotzky,Starace,Goreslavski,Popruzhenko,Popruzhenko1}. 
\begin{figure}
\includegraphics[width=0.5\textwidth]{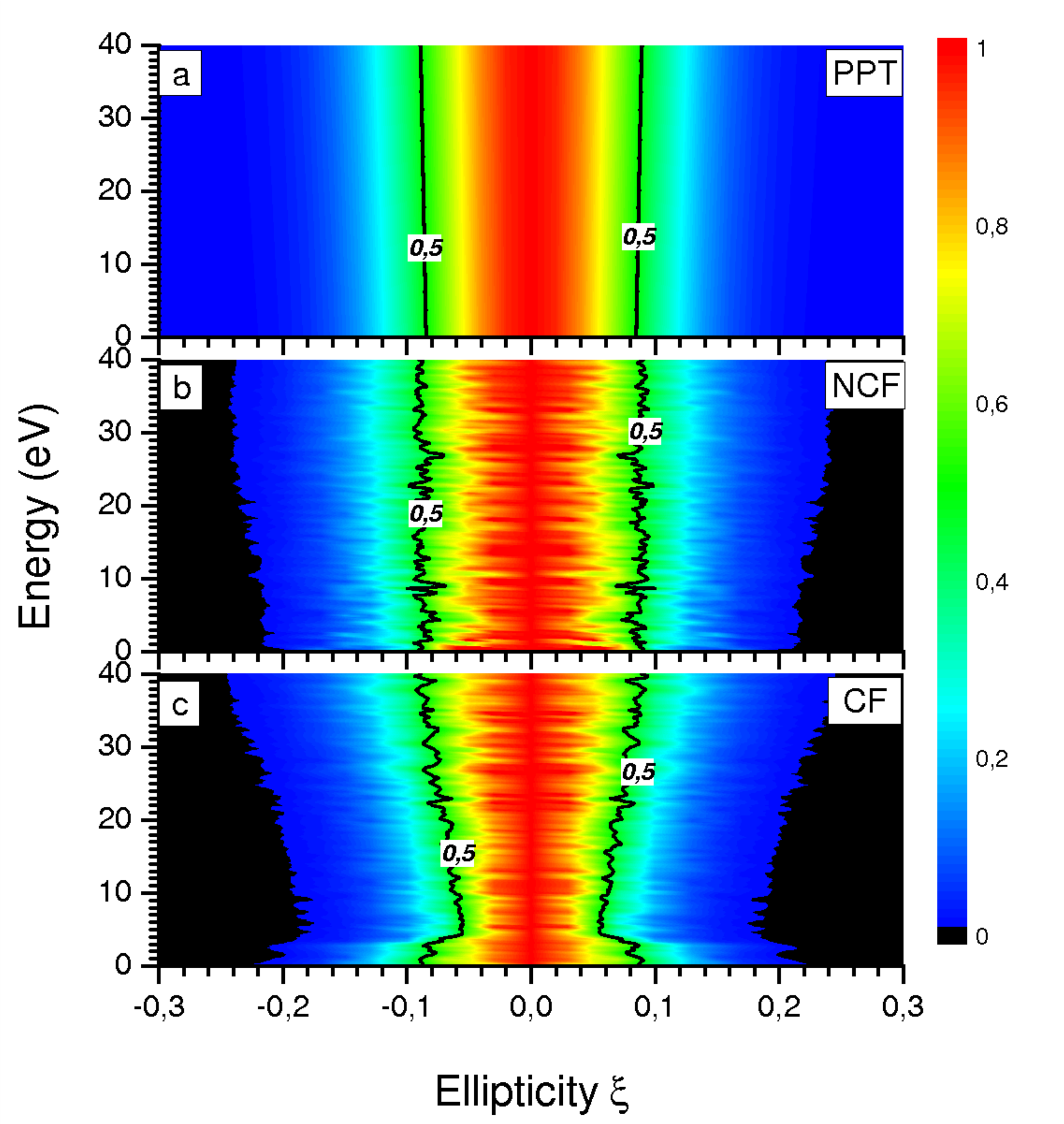}
\caption{(color online) The photoelectron normalized yield emitted along m.p.a. versus ellipticity and photoelectron energy: (a) Analytical estimate of the yield in the PPT model via Eq. (\ref{dW}); (b) Numerical simulations without taking into account Coulomb field effects;  (c) Numerical simulations with  Coulomb field effects. The laser intensity is $I_0=9\times 10^{13}$ W/cm$^2$ and wavelength $\lambda=2 \,\mu $m. The target atom is hydrogen. The curves with value 0.5 are for eye guide. }
\label{totalangle}
\end{figure}

In this paper, we investigate how the Coulomb field effects of the atomic core modify the dodging phenomenon in above-threshold ionization in an elliptically polarized  mid-infrared laser field. First, our attention is focused on the ellipticity and energy resolved photoelectron yield along m.p.a. up to moderate ellipticities. We show that due to CF, a remarkable energy dependence arises in the yield. What is more surprising is the energy dependence of the yield for the low-energy domain [decreasing of the yield with increasing energy, in the energy interval of $(0,4)$ eV in Fig. \ref{totalangle}] is reversed with respect to that for the high-energy domain [in the energy interval of $(4,40)$ eV in Fig. \ref{totalangle}]. We demonstrate a direct relationship of this peculiar energy dependence of the yield with the LES appearance and with the specific features of multiple forward scattering of the ionized electron by the atomic core. Secondly, we investigate the role of the Coulomb field effects at high ellipticities of the field. The CF is shown to enhance the photoelectron total yield along m.p.a. at high ellipticities. Moreover, the enhancement factor peaks at an intermediate value of ellipticity.

We employ a semi-classical model incorporating tunneling and Coulomb field effects. Note that the CF effect and LES are conspicuous in mid-infrared laser fields when the Keldysh parameter is small,
$\gamma=\sqrt{I_p/2U_p}\ll 1$ \cite{Becker_review}, where $I_p$ is the ionization potential  and  $U_p$ the ponderomotive energy. In this case the electron dynamics after tunneling is mainly classical and not disguised by quantum interference. 

The structure of the paper is as follows. In Sec. \ref{model} our theoretical model is presented. Coulomb focusing up to moderate ellipticity values is considered in Sec. \ref{low_ellipt}. In this section the appearance of  LES and peculiar energy scaling of the normalized yield are shown and discussed. Sec. \ref{large_ellipt} is devoted to the discussion of the Coulomb focusing effect at high ellipticities of the laser field. Sec. \ref{conclusion} concludes our discussion.

\section{Theoretical model}\label{model}

Our investigation is based on the classical-trajectory Monte Carlo method, with tunneling and the Coulomb field of the atomic core fully taken into account.
The ionized electron wave packet is formed according to the Perelomov, Popov, Terent'ev (PPT) ionization rate \cite{PPT,ADK} which is further propagated by the classical equations of motion:
\begin{eqnarray}
 \frac{d \textbf{p}}{dt}=-\textbf{E}(t)+\mathbf{\nabla} V_C(r),
\label{equation-of-motion}
\end{eqnarray}
where $V_C(r)=Z/r$ is the Coulomb potential of the atomic core [the target atom is hydrogen, $Z=1$, and atomic units are used throughout the paper]. The laser field $\textbf{E}(t)=(E_{x}(t),E_{y}(t),0)$ is elliptically polarized:
\begin{eqnarray}
E_{x}(t)&=& E_{0} f(t) \cos \omega t, \nonumber \\ 
E_{y}(t) &=& - \xi E_{0} f(t) \sin \omega t,
\label{field}
\end{eqnarray}
with the ellipticity $\xi$ ($|\xi| \leq 1$).
The electrons are born at the tunnel exit with the coordinates
\begin{eqnarray}
x_{i} = x'_{i}\cos \beta, \,\,\,\,\,\,
y_{i} = x'_{i}\sin \beta, \,\,\,\,\,\,
z_{i} = 0,
\label{coordinate}
\end{eqnarray}
where $\beta = \arctan [-\xi \tan \varphi_i ]$, $\varphi_i\equiv \omega t_{i}$ is the ionization phase and $x'_{i}$ the initial position along the laser polarization direction \cite{Landau77}.
The initial momentum components are
\begin{eqnarray}
p_{ix} &=& -p_{i\bot} \cos\alpha \sin\beta,\nonumber \\ 
p_{iy} &=& p_{i\bot} \cos\alpha \cos\beta,\nonumber \\  
p_{iz} &=& p_{i\bot} \sin\alpha,
\label{momentum}
\end{eqnarray}
where $\alpha$ is the angle between $p_{i\bot}$ and the axis $y^{\prime}$ [the $y$ axis after rotation by an angle $\beta$ around axis $z$] which is randomly distributed within
the interval $(0, 2\pi)$. The transverse momentum $p_{i\bot}$ follows the
corresponding PPT distribution \cite{PPT}. The positions and momenta of
electrons after interaction with the laser pulse are used to calculate the final
asymptotic momenta \cite{Landau_Mechanics} at the detector. Only
electrons emitted along the $x$ direction (m.p.a.) within an opening angle $\theta_0=\pm 2.5^{\circ}$ are collected. The laser pulse profile is half-trapezoidal, constant for the first ten cycles and ramped off within the last three cycles. The number of propagated electrons is $ 10^6 $ and the convergence is checked via double increase of the electron number. The electrons are launched within the first half cycle ($\omega t_{i}\in [0,\pi]$), since there are no multi-cycle interference effects in the classical theory.
The model has been confirmed to provide an adequate description for the strong
field dynamics in the mid-infrared regime \cite{Liu}.

\section{Low-energy structure}\label{low_ellipt}

\begin{figure}[t]
\includegraphics[width=0.5\textwidth]{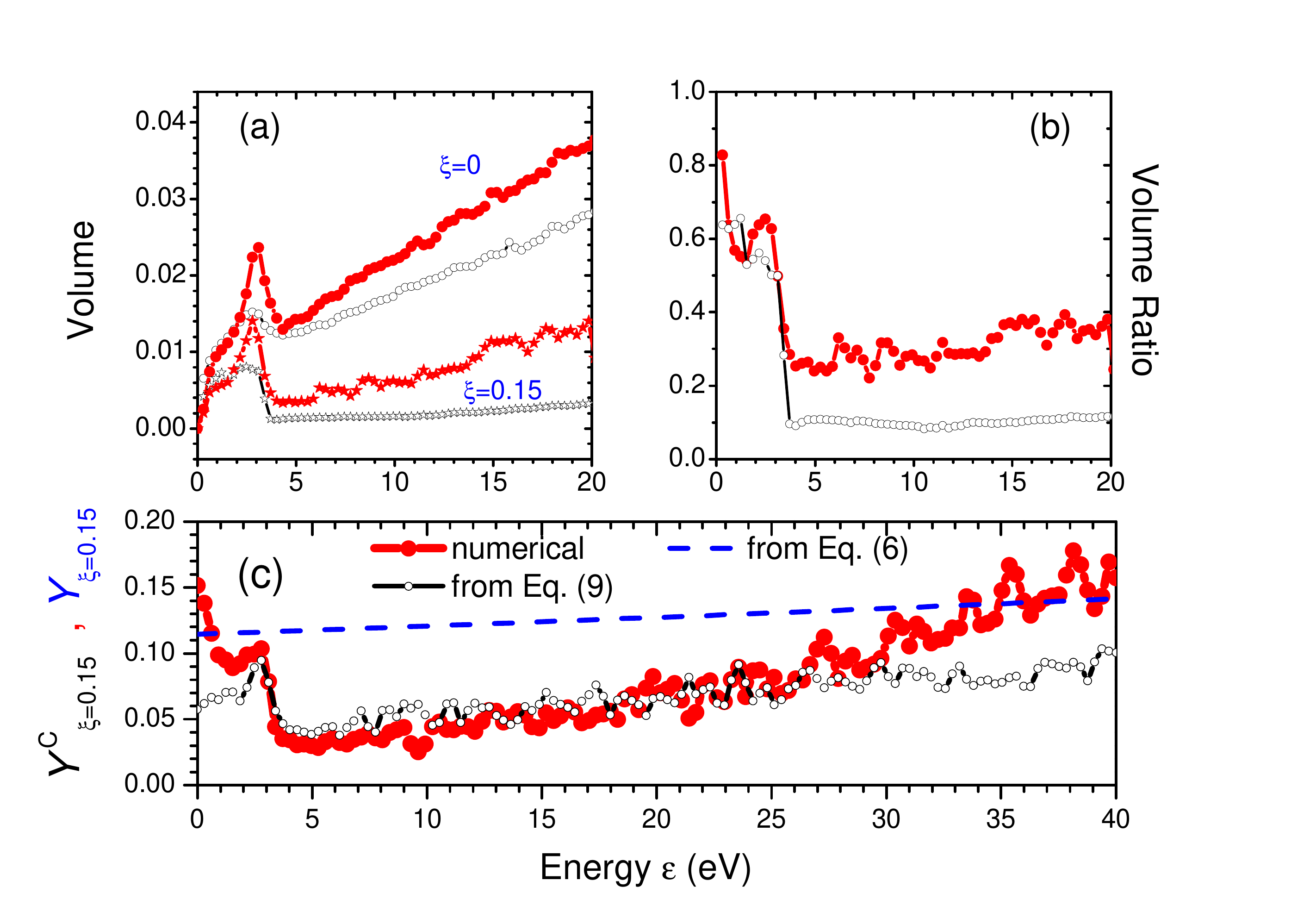}
\caption{(color online) (a) The initial phase-space volume of contributing electrons $d^2\textbf{p}^C_{i\bot \xi}$ versus photoelectron energy at $\xi=0$ and $\xi=0.15$, numerical calculation (filled circles) and estimation (see text) (open circles); (b) The  ratio of the initial momentum-space volumes $d^2\textbf{p}^C_{i\bot \xi}/d^2\textbf{p}^C_{i\bot 0}$, numerical (filled circles) and estimated (open circles); 
(c) The normalized yield, numerical (filled circles) and estimated via Eq. (\ref{Y}) (open circles) [is obtained by multiplying the numerical values of the volume ratio from Fig. \ref{Yield} (b) by $Y_\xi(\varepsilon)$ of Eq. (\ref{dW})]. The dashed line shows $Y_\xi(\varepsilon)$ via Eq. (\ref{dW}). The laser and atom parameters are the same as in Fig. \ref{totalangle}. 
}
\label{Yield}
\end{figure}

As we are concerned with the CF impact on the dodging phenomenon, let us first recall how this phenomenon arises when the Coulomb field effect is neglected \cite{Paulus_98}. The electrons born near the maximum of the elliptically polarized field tend to drift
towards the minor axis of the polarization ellipse. Consequently, the electron should have a rather large transverse momentum $|p_y|\approx |A_y(\varphi_i)|\approx \xi E_0/\omega$ at the tunnel exit to counteract the drift and to reach the final state where the momentum points along  m.p.a. (along $\hat{\textbf{x}}$). Here $A_y(\varphi_i)$ is the field vector-potential component at the ionization phase $\varphi_i$. 
As a result, the ionization probability for electrons moving along m.p.a. is exponentially decreased with rising ellipticity (up to $\xi\lesssim \sqrt{E_0}$): 
\begin{eqnarray}
W_{\xi}\propto \exp(-\xi^2 E_0^2/\omega^2\Delta_{\bot}^2),
\label{Wxi}
\end{eqnarray}
where $\Delta_{\bot}^2\equiv E_0/\sqrt{2I_p}$ is the width of the transverse  momentum distribution \cite{Usp}.
In describing the dodging phenomenon \cite{Paulus_98}, the photoelectron yield $Y_{\xi}(\varepsilon)$ along m.p.a., normalized to the yield in a linearly polarized field, has been introduced and measured, where $\xi$ and $\varepsilon$ indicate the ellipticity  and  electron energy dependence of the yield, respectively.
Using the PPT  probabilities for the photoelectron momentum distribution in an elliptically polarized field $d^3W_{\xi}/d^3\textbf{p}$ \cite{Usp,Mur}, where the CF effects for the electron dynamics in the continuum are neglected,  the normalized yield can be derived: 
\begin{eqnarray}
Y_{\xi}(\varepsilon)\equiv \frac{\int_{\Delta \Omega}\left(d^3W_{\xi}/d^3\textbf{p}\right)d\Omega}{\int_{\Delta \Omega}\left(d^3W_{0}/d^3\textbf{p} \right)d\Omega}.
\label{Yxi}
\end{eqnarray}
Here the integration is carried out over the solid angle $\Delta \Omega$ along the m.p.a. with an opening angle $\theta_0$. At $\xi\lesssim \sqrt{E_0}$ the yield reads 
\begin{eqnarray}\label{dW}
Y_{\xi}(\varepsilon)&=&\exp \left\{-\xi^2\left(\frac{E_0^2}{\omega^2\Delta_{\bot}^2}- \frac{2\varepsilon}{\Delta_{\parallel}^2}\right)\right\}\frac{2I_1\left( \frac{2\sqrt{2\varepsilon}\xi E_0\theta_0}{\omega\Delta_{\bot}^2}\right)}{\frac{2\sqrt{2\varepsilon}\xi E_0\theta_0}{\omega\Delta_{\bot}^2}} 
\end{eqnarray}
where  $\Delta_{ \parallel}^2\equiv 3E_0^3/(2I_p)^{3/2}\omega^2$ is the width of the  longitudinal momentum distribution and $I_1(Z)$ is a modified Bessel function.  As Eq. (\ref{dW}) shows, the yield exponentially decreases with rising ellipticity $\xi$ and depends weakly on energy $\varepsilon$, when the CF effects are neglected, see  Fig. \ref{totalangle} (a, b). The cut across Fig. \ref{totalangle} at a fixed energy 
illustrates the first point, while the cut across Fig. \ref{totalangle} at a fixed ellipticity  illustrates the second point [see also the dashed line in Fig. \ref{Yield} (c) below].  
With the Coulomb field included, our numerical simulation shows an anomalous energy dependence of the 
yield for ellipticities up to $\sim 0.3$, see Fig. \ref{totalangle} (c) and also Fig. \ref{Yield} (c) below: the yield decreases  with increasing energy in the low-energy interval of $(0,4)$ eV, while the energy dependence is reversed in the high-energy domain of $(4,40)$ eV.

To understand the role of the Coulomb field effects, we have calculated the photoelectron normalized yield, neglecting the momentum change induced by Coulomb field, either for the final longitudinal or for the final transverse momentum. This is accomplished by propagating an electron  with certain initial conditions twice: taking into account $V_C(r)$ in Eq. (\ref{equation-of-motion}) or neglecting it. Afterwards, we replace the longitudinal (transverse) component of the exact final momentum with that derived from the calculation neglecting $V_C(r)$. We noted  that neglecting the Coulomb field for the longitudinal momentum has no significant effect on the yield,  while the same for the transverse momentum, eliminates the peculiar energy dependence. As the CF is induced by the reduction of the transverse momentum of the ionized electron \cite{Villeneuve}, we conclude that the CF is behind the anomalous energy dependence of the normalized yield. 
\begin{figure}
\includegraphics[width=0.45\textwidth]{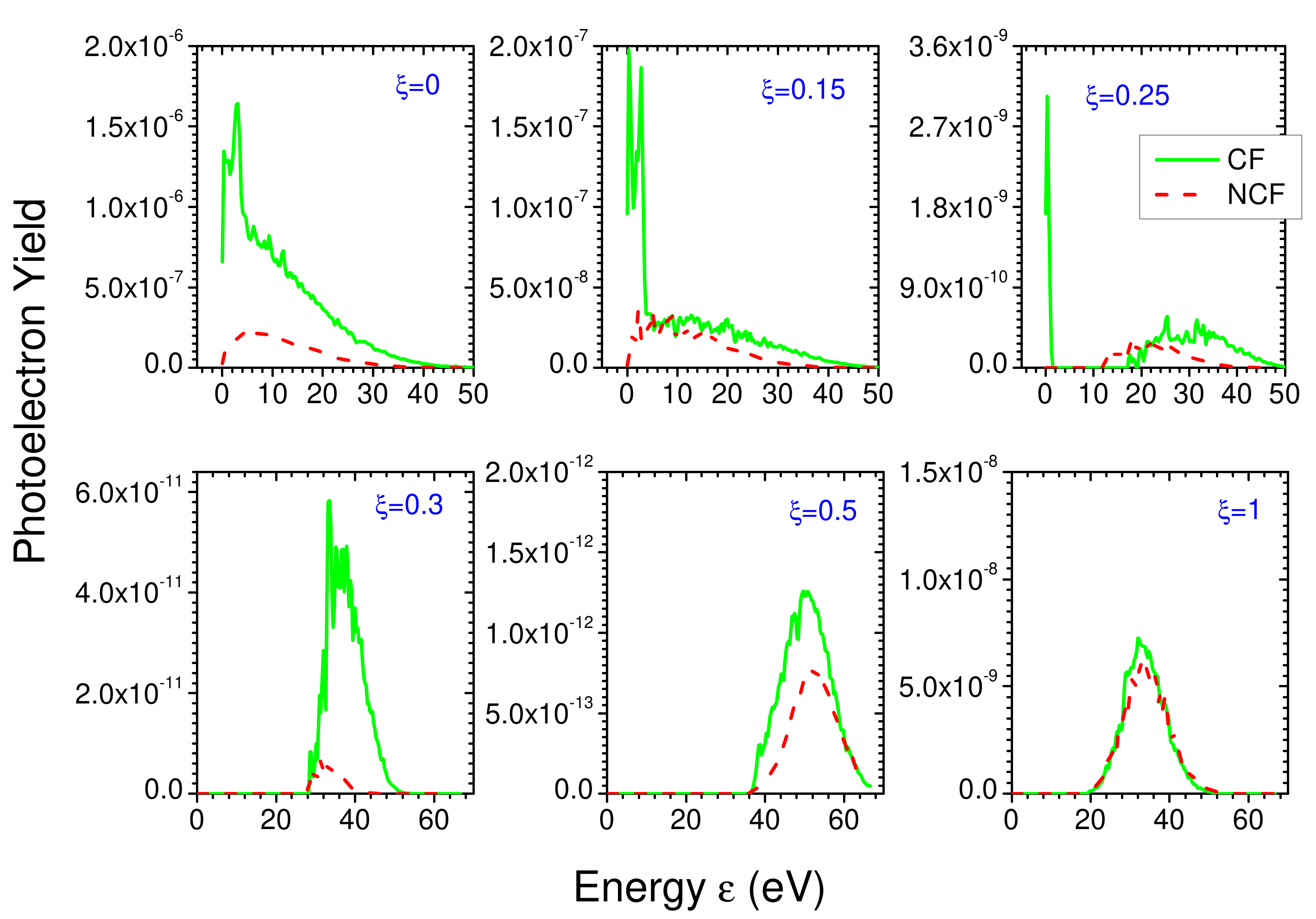}
\caption{(color online) Photoelectron spectra emitted along m.p.a. in an elliptically polarized field with different ellipticity $\xi$, for the cases with CF taken into account (solid) and without CF (dashed). LES is observable up to $\xi\approx 0.25$. The  laser and atom parameters are the same as in Fig. \ref{totalangle}.}\label{LES}
\end{figure}
\begin{figure}
\includegraphics[width=0.45\textwidth]{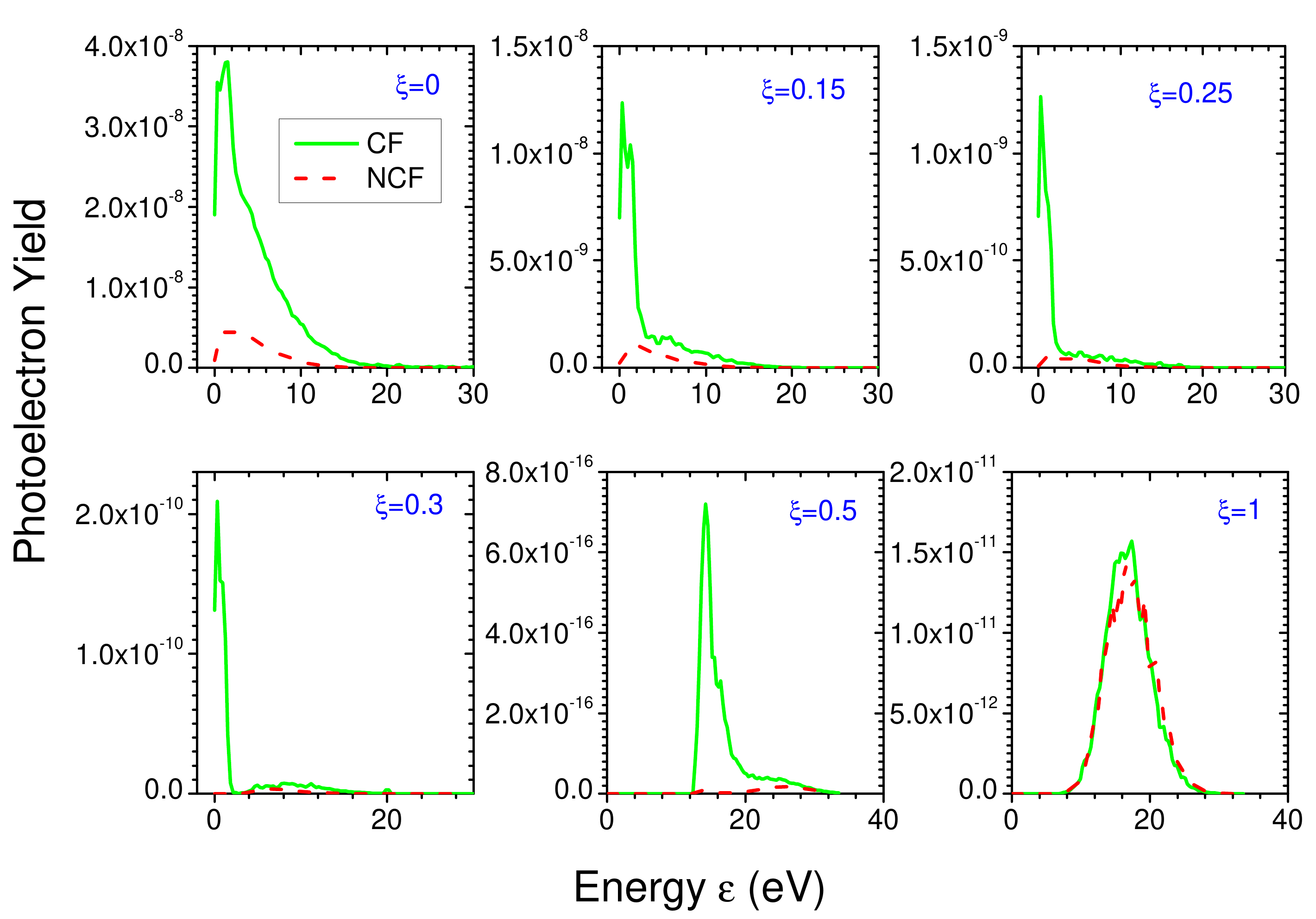}
\caption{(color online) Same as Fig. \ref{LES} with a laser intensity of $I_0=4.5\times 10^{13}$ W/cm$^2$. LES is observable up to $\xi\approx 0.3$. }\label{LES2}
\end{figure}
\begin{figure}
\includegraphics[width=0.45\textwidth]{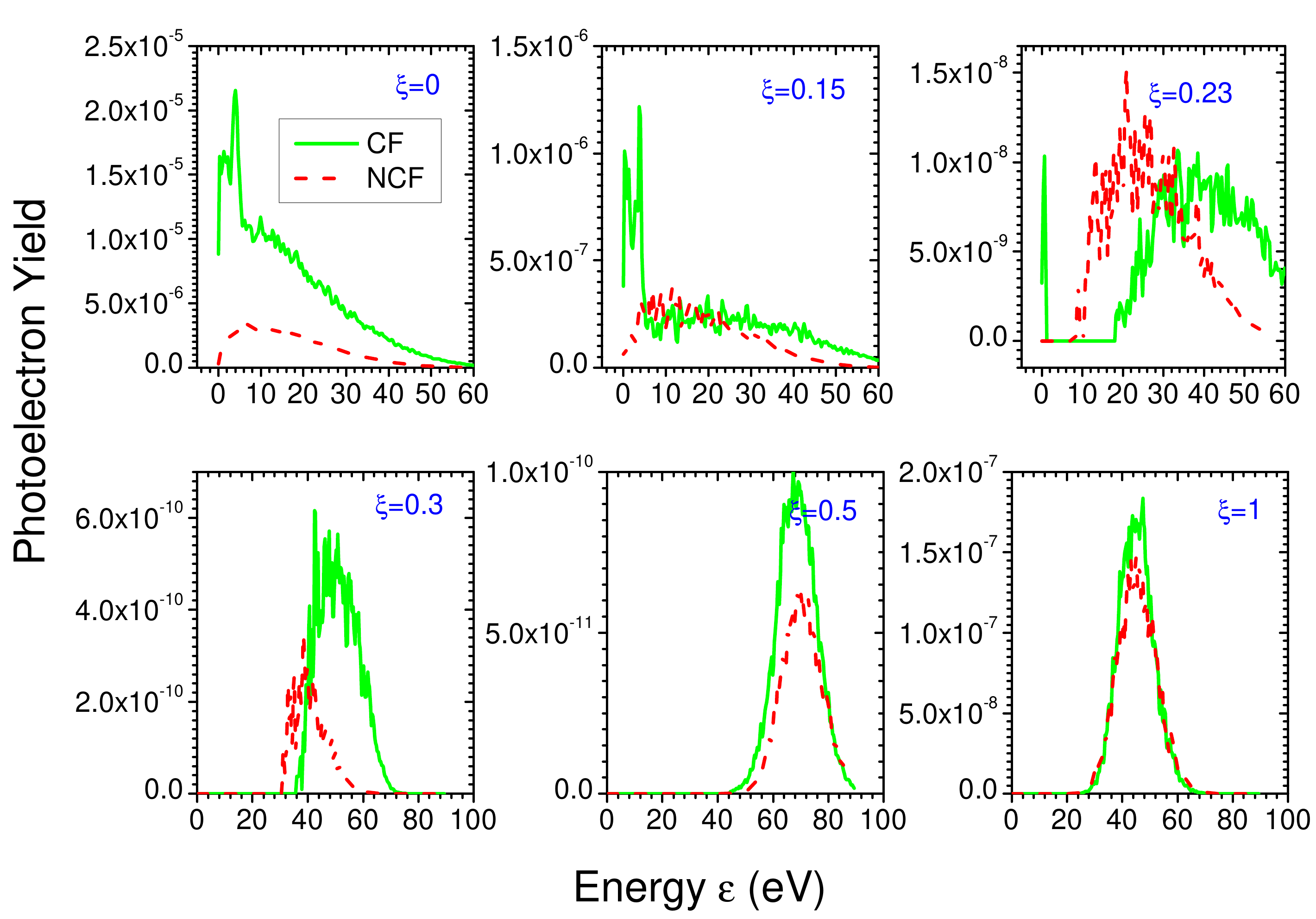}
\caption{(color online) Same as Fig. \ref{LES} with a laser intensity of $I_0=1.2\times 10^{14}$ W/cm$^2$. LES is observable up to $\xi\approx 0.23$. }\label{LES3}
\end{figure}

The  anomalous energy dependence of the  yield is closely connected with the appearance of LES in the energy spectrum of electrons emitted along  m.p.a.. The latter is calculated numerically and shown in Figs. \ref{LES}-\ref{LES3}. We see that the LES [the peak in the energy interval (0, 4) eV] persists up to intermediate ellipticities [$\xi \sim 0.3$ at $I_0=4.5\times 10^{13}$ W/cm$^2$, $\xi \sim 0.25$ at $I_0=9\times 10^{13}$ W/cm$^2$ and $\xi \sim 0.23$ at $I_0=1.2\times 10^{14}$ W/cm$^2$]. Figs. \ref{LES}-\ref{LES3} show also that at larger ellipticities, although there is no LES, the yield is significantly enhanced due to CF. 
The LES and CF due to soft forward scattering persist up to higher values of ellipticity at a given laser intensity than hard recollisions with backward scattering. The upper value of ellipticity for back-scattering ($\xi_{b}$) can be estimated \cite{Shvetsov,Eberly} by equating the lateral drift momentum $\xi E_0/\omega$ to the width of the transverse momentum distribution $\Delta_\bot$: $\xi_{b}\sim \omega/\sqrt{E_0}(2I_p)^{1/4}$. For the parameters chosen in Fig. \ref{LES}, $\xi_{b} \approx 0.098$, while the LES for the same parameters persists up to $\xi\approx 0.3$. Nevertheless, the scaling with a laser intensity of the upper limit of ellipticity for LES is surprisingly well reproduced by the above estimation $\xi\sim  I_0^{-1/4}$.

\begin{figure}[t]
\includegraphics[width=0.5\textwidth]{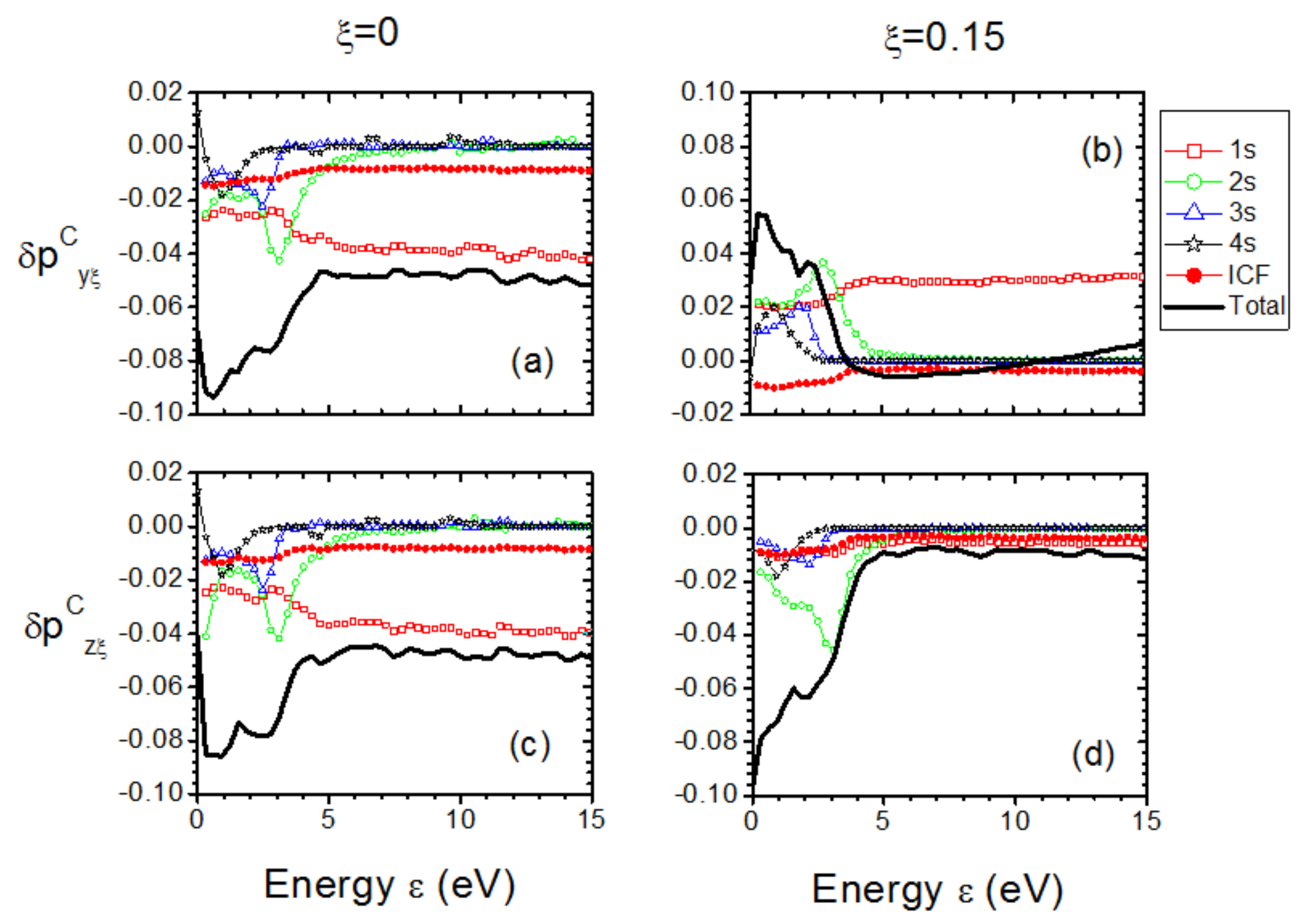}
\caption{(color online)  The transverse momentum change due to CF versus photoelectron energy. The total value according to numerical simulations (black line). The estimated contributions from high-order forward scattering (here up to $4$th) and from initial CF (ICF) at the tunnel exit are shown separately. (a,c) $\xi=0$; (b,d) $\xi=0.15$. The laser  and atom parameters are the same as in Fig. \ref{totalangle}.
}
\label{momentum_change}
\end{figure}

To explain the reason of the  anomalous energy dependence of the normalized yield at moderate ellipticities, we proceed with an estimation of the yield taking into account CF. First, we express the momentum distribution of the ionized electrons at the detector $d^3W_{\xi}/d^3\textbf{p}$ via the distribution function over the transverse momentum $\textbf{p}_{i\bot}$ (with respect to the field at the ionization moment) and the ionization phase $\varphi_i$ at the tunnel exit:
\begin{eqnarray}
\label{d3W}
\frac{d^3W_{\xi}}{d^3\textbf{p}}=\frac{d^3W_{\xi}(\varphi_i,\textbf{p}_{i\bot})}{d\varphi_id^2\textbf{p}_{i\bot}}J_{\xi}, 
\end{eqnarray}
where $J_{\xi}\equiv \frac{\partial (\varphi_i,\textbf{p}_{i\bot})}{\partial(p_{\parallel},\textbf{p}_{\bot })}$ is the transformation Jacobian. The final momentum of the electron $\textbf{p}$ is a function of the variables  at the tunnel exit $\textbf{p}=\textbf{p}_\xi(\varphi_i,\textbf{p}_{i\bot})$ which depends on $\xi$ and is essentially disturbed by CF. Consequently, the dependence of the distribution function  $d^3W_{\xi}(\varphi_i(\textbf{p}),\textbf{p}_{i\bot }(\textbf{p}))/d\varphi_id^2\textbf{p}_{i\bot}$ on the final momentum is also modified due to CF. Fortunately, 
this modification is negligible when the transverse momentum induced by CF $\delta p_{\bot }^C$ is rather small: $\delta p_{\bot}^C \ll \omega/\xi \sqrt{2I_p}$ [derivation of this condition is given in Sec. \ref{large_ellipt}, see Eq. (\ref{PyC-condition}) below]. This is the case in the region of the anomalous energy dependence of the yield, see e.g. Fig. \ref{momentum_change} (b,d). Thus, in this region the CF effect is mainly contained in the Jacobian $J_{\xi}$. With this approximation, the normalized yield with CF effects $Y_\xi^C(\varepsilon)$ can be expressed via the yield without CF $Y_\xi(\varepsilon)$ given by Eq. (\ref{dW}): 
$Y_\xi^C(\varepsilon)\approx Y_\xi(\varepsilon)(J^C_\xi/J^C_0)(J_{0}/J_\xi)$,
where the index  ``C'' indicates  incorporation of the CF effects. 
The ratio of the Jacobians can be expressed by the ratio of the initial momentum-space at the tunnel exit at a fixed momentum-space at the detector: 
\begin{eqnarray}
\label{YC=YP}
\frac{J^C_\xi}{J^C_0}\approx \frac{d^2\textbf{p}^C_{i\bot \xi}}{d^2\textbf{p}^C_{i\bot 0}}. 
\end{eqnarray}
Moreover, without CF effects $ J_{\xi}/J_0\approx d^2\textbf{p}_{i\bot \xi}/d^2\textbf{p}_{i\bot 0}=1$, because when CF is neglected, the final momentum-space at the detector equals that for the contributing electrons at the tunnel exit for any ellipticity. In conclusion, the yield with CF effects $Y_{\xi}^C(\varepsilon)$ can be expressed via the yield without CF $Y_{\xi}(\varepsilon)$ and the ratio of the initial momentum-space volumes $d^2\textbf{p}^C_{i\bot \xi}/d^2\textbf{p}^C_{i\bot 0}$ at a fixed final momentum-space:
\begin{equation}
Y_{\xi}^C(\varepsilon)=\frac{d^2\textbf{p}^C_{i\bot \xi}}{d^2\textbf{p}^C_{i\bot 0}} \,\,\,Y_{\xi}(\varepsilon).
\label{Y}
\end{equation}
\begin{figure}[b]
\includegraphics[width=0.5\textwidth]{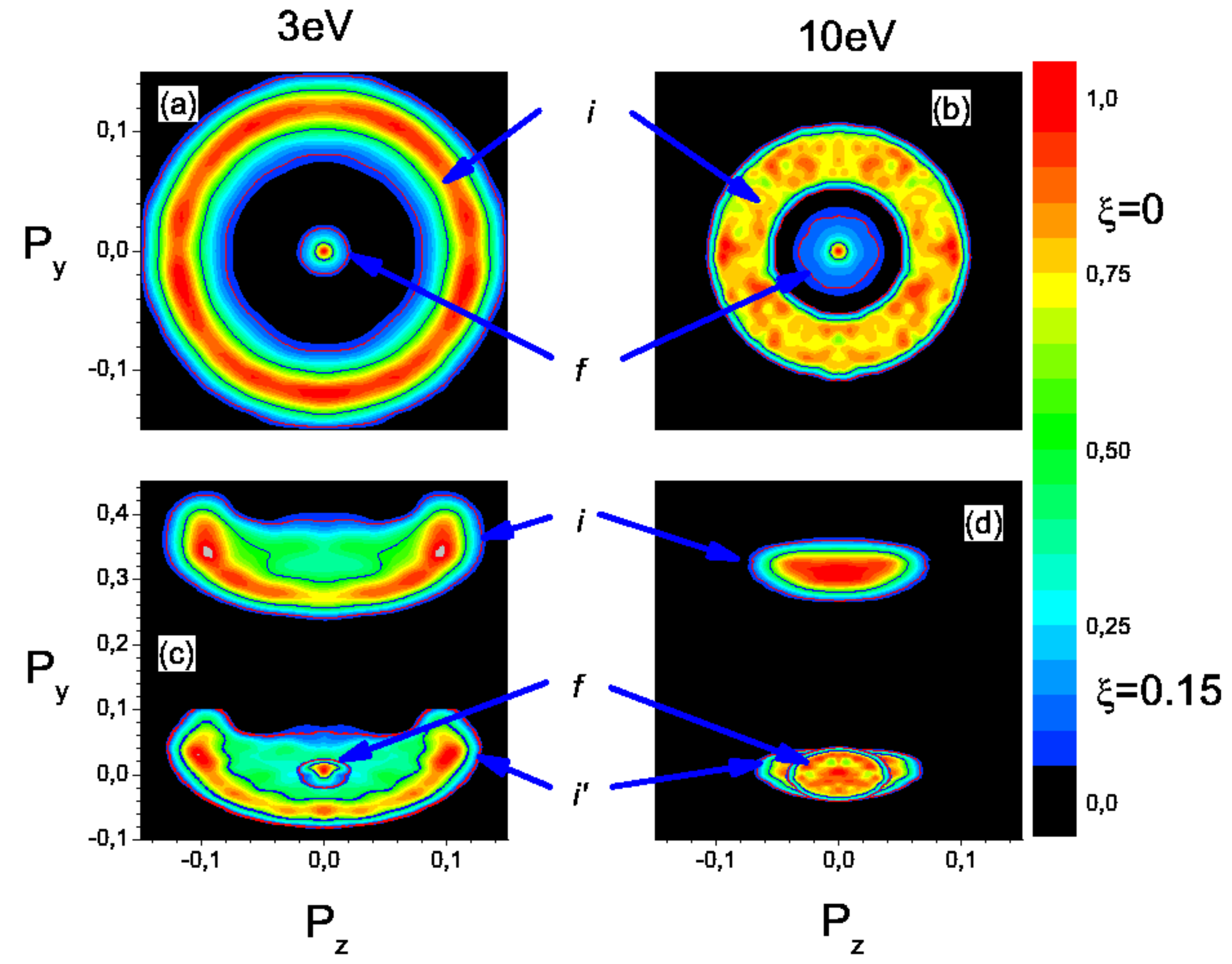}
\includegraphics[width=0.2\textwidth]{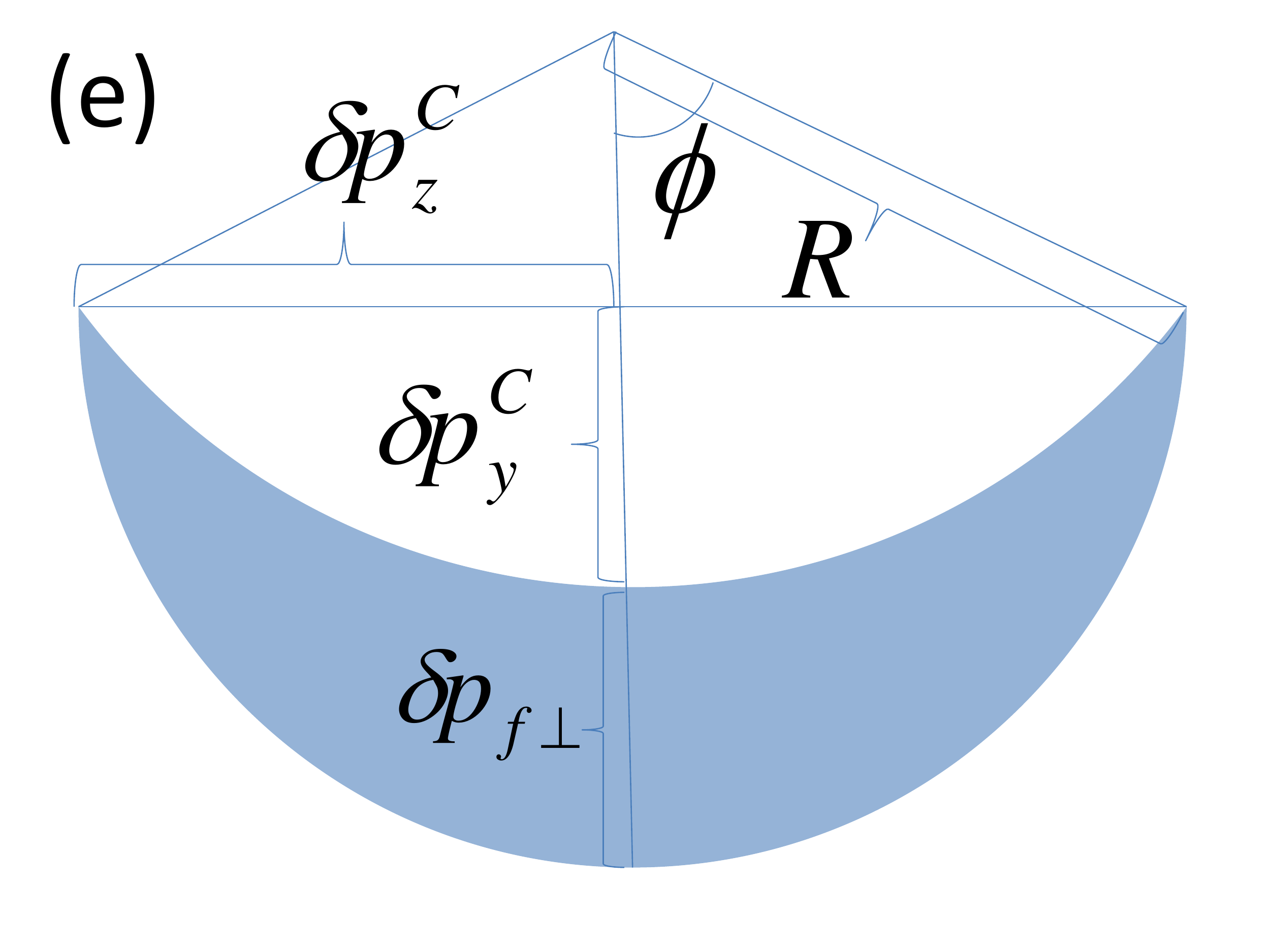}
\caption{(color online) The initial [at the tunnel exit, marked by ``i''] and the final [at the detector, marked by ``f''] transverse momentum distributions of electrons with specified energies, (a,c) $\varepsilon=3$ eV and  (b,d) $\varepsilon=10$ eV  for the cases of linear  (a,b) or elliptical polarization ($\xi=0.15$) (c,d), respectively. The initial distribution shifted by $A_y(\varphi_i)$ is indicated as ``$i^{\prime}$''; (e) To the estimation of the momentum-space volume, see text Eq. (\ref{d2pCxi}). The laser and atom parameters are the same as in Fig. \ref{totalangle}. }
\label{phase_space}
\end{figure}
We have calculated numerically the  ratio of the momentum-space volumes $d^2\textbf{p}^C_{i\bot \xi}/d^2\textbf{p}^C_{i\bot 0}$, see Fig. \ref{Yield} (b), and checked the validity of the estimation of Eq. (\ref{Y}) [open circles in Fig. \ref{Yield} (c)] comparing it with the exact numerical calculations [filled circles in Fig. \ref{Yield} (c)]. 
One can see that in the energy interval of $(2,30)$ eV, which includes the region of peculiar behavior of $Y_{\xi}^C(\varepsilon)$, the above estimation  describes the  energy dependence of the yield appropriately.
As $Y_\xi(\varepsilon)$ without CF [dashed line in Fig. \ref{Yield} (c)] depends on energy monotonously, we can conclude that the minimum at $\varepsilon\approx 4$ eV in the yield $Y_{\xi}^C(\varepsilon)$ with CF originates from the ratio of the momentum-space volumes of contributing electrons at the tunnel exit and is due to CF influence. 

Now, let us see why the  ratio of the momentum-space volumes has a minimum. We can estimate the initial transverse momentum-space volume at the tunnel exit for the electrons which at the detector move along m.p.a., with a given energy interval as follows. In the case of linear polarization, the electron gets a transverse momentum kick $\delta p_{\bot}^C$ due to CF during its motion from the tunnel exit to the detector, therefore, only the electrons which are distributed at the tunnel exit on a ring in a transverse momentum-space with a radius $\delta p_{\bot}^C$ will move along m.p.a. at the detector, see Fig. \ref{phase_space} (a,b). The width of the ring is of the order of the final transverse momentum allowance $\delta p_{f\bot}\sim p\theta_0$, with $p=\sqrt{2\varepsilon}$. Therefore, in the linear polarization case the initial momentum-space of contributing electrons is proportional to the transverse momentum change due to CF 
\begin{equation}
d^2 \textbf{p}^C_{i\bot 0}\sim 2\pi p\theta_0 \delta p_{\bot}^C. 
\label{d2pC}
\end{equation}
In an elliptically polarized laser field, the momentum-space distortion is more complex, see Fig. \ref{phase_space} (c,d). 
In this case, the transverse phase-space of contributing electrons at the tunnel exit can be expressed via the components of the momentum change $\delta p_{y,z \xi}^C$ induced by CF, see Fig. \ref{phase_space} (e): 
$d^2 \textbf{p}^C_{i\bot \xi}\sim 2R\phi \delta p_{f\bot}$, with $R=[(\delta p_{y \xi}^C)^2+(\delta p_{z \xi}^C)^2]/\delta p_{y\xi}^C$ and $\sin \phi=\delta p_{z \xi}^C/R$, yielding
\begin{equation}
d^2 \textbf{p}^C_{i\bot \xi}\sim \delta p_{f\bot}\frac{(\delta p_{y \xi}^C)^2+(\delta p_{z \xi}^C)^2}{\delta p_{y\xi}^C}\sin^{-1}\frac{|\delta p_{y \xi}^C\delta p_{z \xi}^C|}{(\delta p_{y \xi}^C)^2+(\delta p_{z\xi}^C)^2}.
\label{d2pCxi}
\end{equation}
In the most essential range of the photoelectron energy, see Fig. \ref{momentum_change}, $\delta p_{y\xi}^C\ll \delta p_{z\xi}^C$ which allows simplification of the estimate $d^2 \textbf{p}^C_{i\bot \xi}/d^2 \textbf{p}^C_{i\bot 0}\propto \delta p_{y \xi}^C/\delta p_{\bot 0}^C $. These estimations are in qualitative agreement with the exact calculations, as Figs. \ref{Yield} (a,b) show. 
Therefore, the minimum in the ratio of the initial phase-space volumes $d^2 \textbf{p}^C_{i\bot \xi}/d^2 \textbf{p}^C_{i\bot 0}$ is connected with the minimum of the transverse momentum change due to CF $\delta p_{\bot \xi}^C$. The latter for the cases of linear and elliptical polarizations are calculated and shown in Fig. \ref{momentum_change} using the method described in \cite{Liu_JPB}. The partial contributions of high-order scattering events into the total momentum change are shown separately. The total momentum change due to CF, in both cases, exhibits a pronounced minimum at the energy corresponding to the minimum of the ratio of the momentum-space volumes, see Fig. \ref{Yield} (b). This minimum marks the LES region and corresponds to the threshold of the multiple forward scattering. The ionized electrons with final energies larger than 4 eV re-scatter by the atomic core only once, while the electrons with smaller energies can re-scatter several times, inducing larger momentum change, see Fig. \ref{momentum_change}. It is remarkable that contribution of the second forward scattering is more pronounced in the case of elliptical polarization than in the linear one, noted also in \cite{Shvetsov}. The high-order forward scatterings have an essential contribution to the transverse momentum change of the low energy photoelectrons. They are responsible for the creation of the minimum in the  ratio of the momentum-space volumes and, consequently, for the creation of the minimum in the normalized yield at photoelectron low energies. An important conclusion is that the multiple forward scatterings play a significant role in CF also in the elliptically polarized field. It is the reason why the LES persists up to moderate ellipticities.

\section{Coulomb focusing at large ellipticity}\label{large_ellipt}

\begin{figure}
\includegraphics[width=0.5\textwidth]{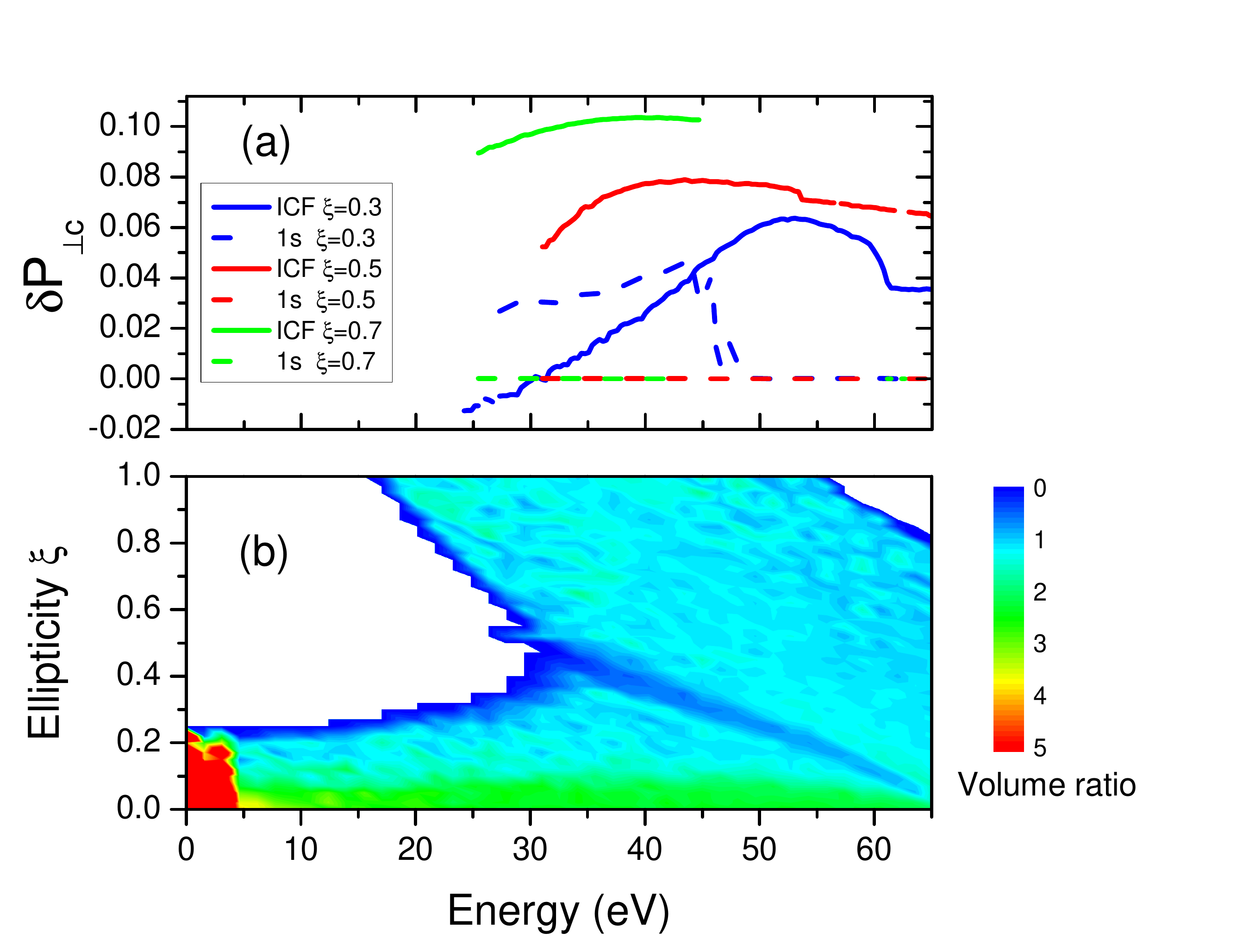}
\caption{(color online) 
(a) The transverse  momentum change due to initial CF and the first forward scattering for large ellipticity values indicated in the inset. (b) The ratio of the momentum-space volumes $d^2\textbf{p}^C_{i\bot \xi}/d^2\textbf{p}^C_{i\bot 0}$  versus photoelectron energy and ellipticity.The laser and atom parameters are the same as in Fig. \ref{totalangle}.
}
\label{Largexi_p}
\end{figure}

At large ellipticities $\xi\gtrsim \sqrt{E_0}$, the character of CF is essentially different. In this domain, the electron drift momentum is dominated by the vector-potential at the ionization moment, with a minor contribution from the initial transverse momentum \cite{Usp}. The contribution of the high-order forward scattering in CF, as Fig. \ref{Largexi_p}(a) shows, becomes monotonous in $\varphi_i$ and perturbative. 
Moreover, the contribution of the forward scattering in CF becomes negligible relative to that of the initial Coulomb disturbance. The momentum transfer at the initial Coulomb disturbance is directed along the field at the ionization moment \cite{Shvetsov-Shilovski}, i.e. it is perpendicular to the drift momentum, causing rotation of the initial momentum distribution around the axis perpendicular to the polarization plane, see also \cite{Shvetsov}. Therefore, the momentum-space volume  will not be essentially modified by CF in this case. In fact, as one can see in Fig. \ref{Largexi_p} (b) the ratio of the initial to final transverse momentum-space volumes is approximately one at large ellipticities $\xi \gtrsim 0.25$.

\begin{figure}[b]
\includegraphics[width=0.4\textwidth]{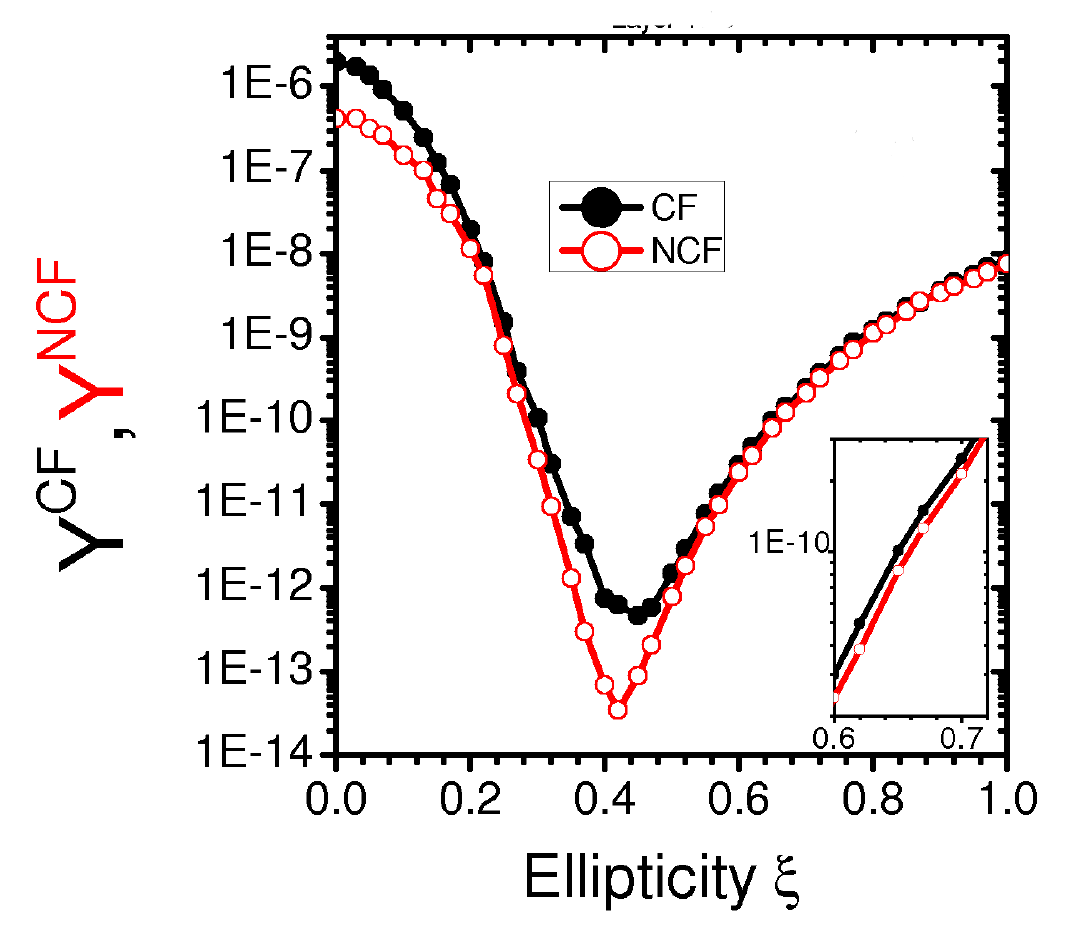}
\caption{(color online) 
The photoelectron normalized yield along m.p.a. integrated over the total energy range for the cases with CF ($Y_{\xi}^{CF}$, black, filled circles)  and  without CF ($Y_{\xi}^{NCF}$, red, open circles).
The laser and atom parameters are the same as in Fig. \ref{totalangle}.
}
\label{Largexi_Y}
\end{figure}

\begin{figure}
\includegraphics[width=0.5\textwidth]{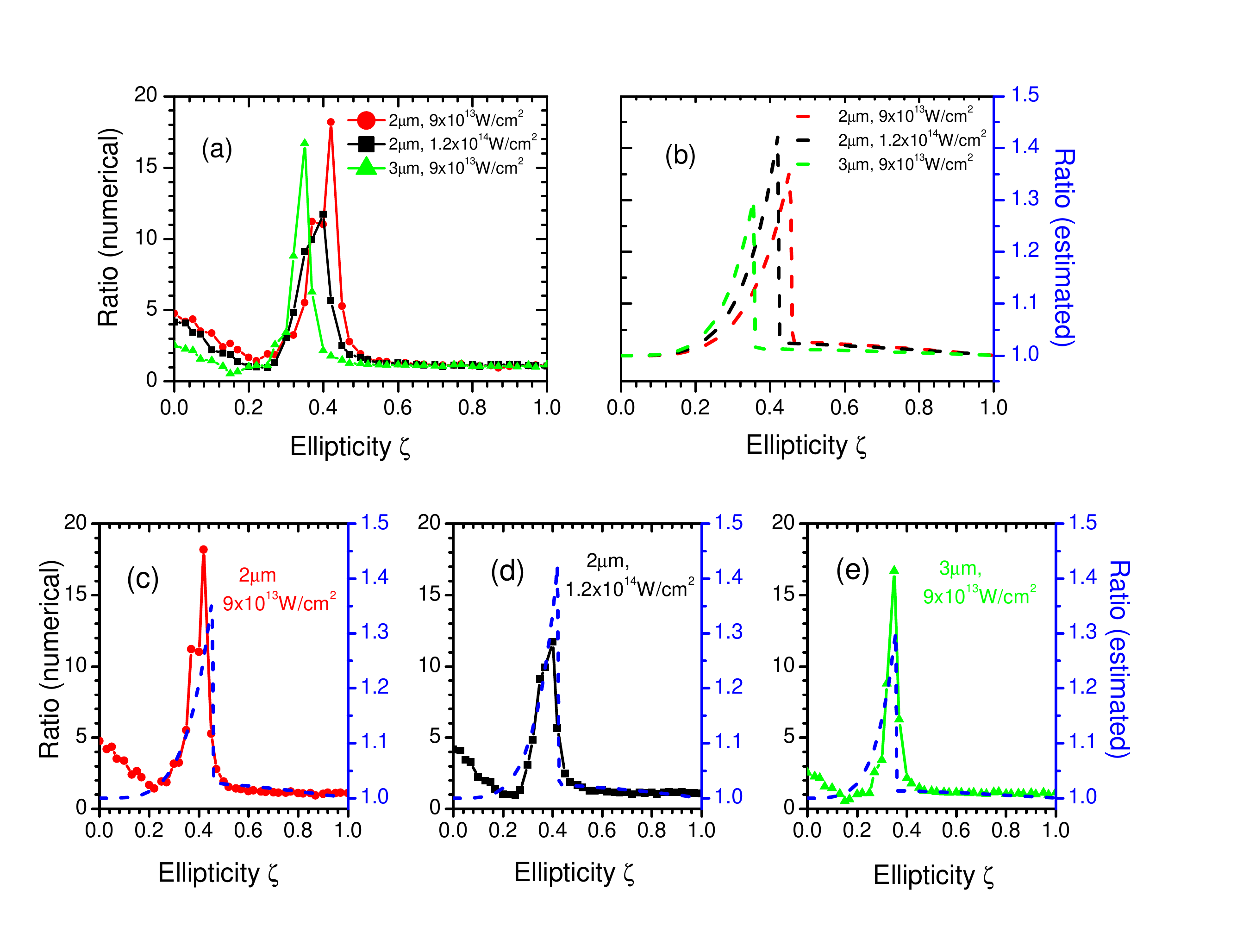}
\caption{(color online) 
The ratio $\rm Y^{CF}/Y^{NCF}$ for different laser intensities ($I_0$) and wavelengths ($\lambda$): (a) numerical simulations, (b) analytical estimations,  according to Eq. (\ref{w_ixi}), (c) $I_0=9\times 10^{13}$ W/cm$^2$, $\lambda =2 \,\,\mu$m;  (d) $I_0=1.2 \times 10^{14} $ W/cm$^2$, $\lambda =2 \,\,\mu$m and (e) $I_0=9\times 10^{13}$ W/cm$^2$, $\lambda =3 \,\,\mu$m; numerical simulation (solid), analytical estimation (dashed).}
\label{Largexi_R}
\end{figure}

The effect of rotation of the momentum distribution due to CF can be monitored by observing the enhancement of the photoelectron yield along m.p.a.. In other words, we investigate the modification of the dodging phenomenon due to CF. We calculate the total normalized yield along m.p.a., see Fig. \ref{Largexi_Y}, and its enhancement factor due to CF, see Fig. \ref{Largexi_R}. We find that the enhancement factor is sharply peaked at an intermediate ellipticity value $\xi_m \approx 0.4$, see Fig. \ref{Largexi_R}. This value corresponds to the minimum of  the yield along  m.p.a., cf. with Fig. \ref{Largexi_Y} 
\footnote{The value of ellipticity $\xi_m$ corresponding to the  minimum of the normalized yield can be estimated equating the exponential terms of Eqs. (\ref{w_i1}) and (\ref{w_i0}) [without CF correction] which leads to a cubic equation:  $\xi_m^3=\upsilon (1-\xi_m)$, with $\upsilon \equiv \frac{4I_p}{3}\frac{\omega^2}{E_0^2}$. From the perturbative solution of the latter by $\upsilon$, one obtains $\xi_m \sim \upsilon^{1/3}\left (1-\frac{\upsilon^{1/3}}{3}\right )$}.

In the following, we give a simple analytical estimate of the CF impact on the tunneling probability that elucidates the origin of the peak of the yield enhancement factor.
The tunneling probability of an electron can be estimated with exponential accuracy
\begin{equation}
w_i(\varphi_i,p_{i\bot })\propto \exp\left(-2E_a/3E(\varphi_i)-p^2_{i\bot}/\Delta^2_{\bot}\right), 
\label{w_i}
\end{equation}
where $\varphi_i=\omega t_i$ is the laser phase at the tunneling moment, $p_{i \bot}$ the initial transverse momentum with respect to the instantaneous direction of the field, and $E_a=(2I_p)^{3/2}$ the atomic field. The electron momentum at the detector is
\begin{equation}
\textbf{p}=\textbf{p}_i-\textbf{A}(\varphi_i)+\delta \textbf{p}^C,
\label{p}
\end{equation}
where $\textbf{p}_i$ is the electron initial momentum at the tunnel exit, $\textbf{A}(\varphi_i)$ the vector-potential at the ionization moment and $\delta \textbf{p}^C$ the momentum change due to CF
[in the case of large ellipticities, it is due to initial CF].

\begin{figure}[b]
\includegraphics[width=0.35\textwidth]{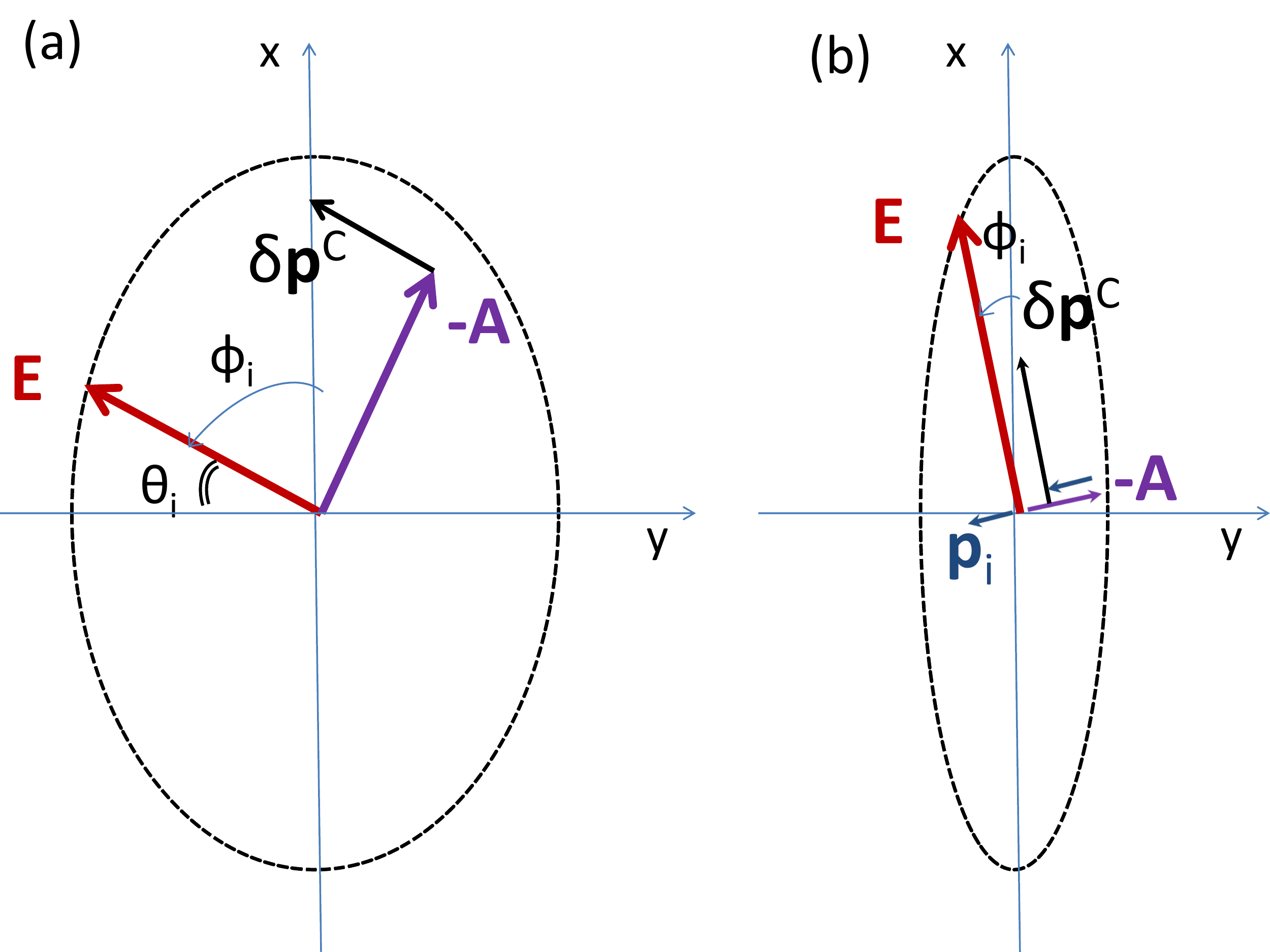}
\caption{(color online) The geometry of the ionization near the  phase of the laser field when the ionization probability is maximal: (a) $\xi \rightarrow 1$, (b) $\xi \rightarrow 0$. x-axis is along m.p.a.}
\label{ellipt1}
\end{figure}

In the limit $\xi\rightarrow 1$, the ionization of electrons which at the detector move along m.p.a. takes place mostly near the laser phase when the vector potential points along  m.p.a., see Fig. \ref{ellipt1} (a). For the field given by Eq. (\ref{field}) the vector-potential is
\begin{equation}
A_x(\varphi_i)=-A_0\sin \varphi_i, \,\,\,\,\, A_y(\varphi_i)=-\xi A_0\cos \varphi_i,
\label{vector-ptential}
\end{equation}
with $A_0=E_0/\omega$, and the ionization phase is close to $\varphi_i \approx \pi/2$. Introducing
$\theta_i\equiv \pi/2-\varphi_i \ll 1$ and expanding the field over a small parameter $\theta_i$ up to the second order, we obtain for the total field 
\begin{equation}
E(\varphi_i)=\sqrt{E_x^2+E_y^2}\approx\xi E_0[1+\theta_i^2(1-\xi^2)/2\xi^2].
\label{E1}
\end{equation}
The ionization phase is determined from vanishing of the final transverse momentum with respect to m.p.a., namely, $p_y=p_{i y}-A_y(\varphi_i)+\delta p_y^C\approx 0$. The probability is maximal at $p_{i y}\approx 0$. Taking into account also that the longitudinal component (with respect to the laser field) of the momentum change due to initial CF is \cite{Shvetsov-Shilovski}
\begin{equation}
\delta p^C_{\parallel} \approx \pi E(\varphi_i)/(2I_p)^{3/2},
\label{PyC}
\end{equation}
we find that $\delta p_y^C\approx \pi \xi E_0/(2I_p)^{3/2}$ and  
$\theta_i\approx \pi\omega/(2I_p)^{3/2}$.
As a result, the ionization probability in the limit $\xi\rightarrow 1$ is
\begin{equation}
w_i|_{\xi\rightarrow 1}\propto \exp\left\{ -\frac{2E_a}{3\xi E_0}+\frac{\pi^2\omega^2(1-\xi^2)}{3E_0\xi^3(2I_p)^{3/2}}\right\}.
\label{w_i1}
\end{equation}
The last term in Eq. (\ref{w_i1}) describes influence of the initial CF on the ionization probability and leads to the probability enhancement. This is because the electrons which move along  m.p.a., are tunneled out at a larger $\theta_i$, consequently, at a stronger laser field, see Eq. (\ref{E1}), when the CF effect is taken into account.

In the opposite limit $\xi\rightarrow 0$, ionization of the electrons which move along  m.p.a. takes place mostly at $ \varphi_i \approx 0$. In fact, the ionization probability at the laser phase $\varphi_i\approx \pi/2$ is negligible, because of weakness of the field at this phase ($E\sim \xi E_0$). Meanwhile, at $ \varphi_i \approx 0$, the field is large ($E\sim E_0$), the laser induced lateral drift momentum is small [as $A_y\sim \xi E_0/\omega$ with $\xi \ll 1$] and can be compensated by the initial transverse momentum, see Fig. \ref{ellipt1} (b). 
The total field  expanded near $ \varphi_i \approx 0$ reads
\begin{equation}
E(\varphi_i)= E_0[1-\varphi_i^2(1-\xi^2)/2]. 
\label{E2}
\end{equation}
In this case, the initial transverse momentum $p_{i\bot }$ is required to have a nonzero value such that $p_{iy}=p_{i\bot }\cos \varphi_i$ would compensate the drift velocity and result in vanishing of the final lateral momentum. The momentum change due to the initial CF is $\delta p_y^C\approx -\pi \xi E_0 \varphi_i/(2I_p)^{3/2}$. 
Vanishing of the final lateral momentum of the electron $p_y\approx 0$, leads to
\begin{equation}
p_{i\bot}\cos \varphi_i
= \frac{\pi \xi E_0}{(2I_p)^2}\sin\varphi_i- \frac{\xi E_0}{\omega}\cos\varphi_i. 
 \label{pboti0}
\end{equation} 
From the latter $p_{i\bot}$ is derived employing an expansion over $\varphi_i$: 
\begin{equation}
p_{i\bot}\approx \frac{\xi E_0}{\omega}\left[\frac{\pi\omega\varphi_i}{(2I_p)^{3/2}}
-1\right]. 
 \label{pboti}
\end{equation} 
Here, the first  term between the brackets is due to the initial CF. The ionization phase $\varphi_i$ is determined by $p_x$, which can be deduced from the x-projection of Eq. (\ref{p})
\begin{equation}
p_x\approx \frac{\pi E_0}{(2I_p)^{3/2}}+\varphi_i\left[\frac{E_0}{\omega}+p_{i\bot}-\frac{2p_{i\bot}E_0}{(2I_p)^{2}} \right].
\end{equation} 
The latter simplifies for $p_x\gg \pi E_0$ [the electron energy larger than 0.3 eV for the chosen paremeters], yielding $\varphi_i\approx p_x\omega/E_0$. Inserting the obtained values for $\varphi_i$ and $p_{i\bot }$ into Eq. (\ref{w_i}), one gets
\begin{eqnarray}
\label{w_i(p_x)}
w_i(p_x)|_{\xi\rightarrow 0}&\sim& \exp\left(-2E_a/3E_0\right)  \label{w_i(p_x)} \\ 
&\times &  \exp\left\{-\frac{(1-\xi^2)p_x^2}{\Delta_{\parallel}^2} 
-\frac{\xi^2}{\Delta^2_{\bot}}\left[\frac{E_0}{\omega}-\frac{\pi\omega p_x}{(2I_p)^{3/2}} \right]^2\right\}.\nonumber
\end{eqnarray} 
Integrating Eq. (\ref{w_i(p_x)}) over $p_x$ and keeping leading terms  in the exponent (up to $\xi^4$) over expansion by a small parameter $\xi$, we derive the ionization probability in the limit  $\xi\rightarrow 0$ with exponential accuracy
\begin{equation}
w_i|_{\xi\rightarrow 0}\propto  \exp\left\{ -\frac{2E_a}{3E_0}-\frac{\xi^2E_0^2}{\omega^2\Delta_{\bot}^2}+\frac{3\pi^2\xi^4 E_0^3}{\omega^2(2I_p)^{7/2}}\right\}.
\label{w_i0}
\end{equation}
The last term in the exponent in Eq. (\ref{w_i0}) stems from the initial CF. It enhances the ionization probability because the initial transverse momentum of the electron, moving finally along m.p.a., is decreased when CF is accounted for, see Eq. (\ref{pboti}).
Using the derived asymptotic expressions for the ionization probability given by Eqs. (\ref{w_i1}) and (\ref{w_i0}), we interpolate the ionization probability for arbitrary $\xi$: 
\begin{equation}
w_i(\xi)\sim \max\{w_i|_{\xi\rightarrow 0},w_i|_{\xi\rightarrow 1}\},
\label{w_ixi}
\end{equation}
and from the latter estimate the ratio of the yield with to the yield without CF, see Fig. \ref{Largexi_R}.  The estimated ratio reproduces correctly the position of the peak. The CF effect is described by the last terms in the exponent in Eqs. (\ref{w_i1}) and (\ref{w_i0}). These terms increase as one moves away from the asymptotic values of $\xi$. Consequently, enhancement of the ionization probability due to CF is largest at an intermediate ellipticity
\footnote{In Fig. \ref{Largexi_R}, in the domain of small ellipticities $\xi <0.25$,  the numerical result for the enhancement factor is larger than the estimated one. It stems from the ratio of the momentum-space volumes $d^2\textbf{p}^C_{i\bot \xi}/d^2\textbf{p}^C_{i\bot 0}$ discussed in Sec. \ref{low_ellipt}, see Eq. (\ref{Y}) and Fig. \ref{Largexi_p} (b), while the estimates of Eqs. (\ref{w_i1}) and (\ref{w_i0}) of this section deal with the CF influence on the tunneling probability at the tunnel exit and do not include the ratio of the momentum-space volumes.}. 
The peak of the yield enhancement shifts to lower ellipticities with increasing intensity or wavelength. It is correctly predicted by our estimation via Eqs. (\ref{w_i1}) and (\ref{w_i0}), see Fig. \ref{Largexi_R}. However, our qualitative estimates are carried out only with exponential accuracy, neglecting the prefactors. For this reason we cannot predict correctly the absolute value of the yield ratio.

Thus, the normalized yield at large ellipticities is enhanced due to the Coulomb field effect [initial CF]. The reason for the enhancement is that the electrons which drift along  m.p.a. have larger tunneling probability when CF is taken into account. This is due to the fact that either [in the case of $\xi\rightarrow 1$]  the electrons are tunneled out at larger laser field values 
or [in the case of $\xi\rightarrow 0$] the electrons are tunneled out with smaller initial transverse momenta when the initial CF is taken into account. At intermediate ellipticities both of the enhancement mechanisms can contribute which induces a sharp peak in the enhancement factor.

Eq. (\ref{w_i(p_x)}) of this section can be used to deduce the condition for neglecting the CF effects in the distribution function  $d^3W_{\xi}(\varphi_i(\textbf{p}),\textbf{p}_{i\bot }(\textbf{p}))/d\varphi_id^2\textbf{p}_{i\bot}$ at small $\xi$. The CF modification is given by the term proportional to $\pi\omega p_x/(2I_p)^{3/2}$ in the exponent of Eq. (\ref{w_i(p_x)}) which equals the momentum change due to CF $\delta p_{\bot }^C$.  The modification is negligible if $\delta p_{\bot }^C\ll \Delta_{\bot}$ and $\delta p_{\bot }^C\xi E_0/\omega \ll \Delta_{\bot}^2$. The second of these conditions is the strongest and reads
\begin{equation}
\delta p_{\bot }^C\ll \frac{\omega}{\xi \sqrt{2I_p}}.
 \label{PyC-condition}
\end{equation} 
For the parameters used, this yields $\delta p_{\bot }^C\ll 0.14$.

\section{Conclusion}\label{conclusion}

We have investigated the role of Coulomb focusing in above-threshold ionization in a mid-infrared laser field of elliptical polarization.  We have shown that multiple forward scattering of the ionized electron by the atomic core has dominated contribution in Coulomb focusing up to moderate ellipticity values.
The multiple forward scattering causes squeezing of the transverse momentum-space volume, which is the main factor influencing the normalized yield at moderate ellipticities, described by Eq. (\ref{Y}). It is responsible for the peculiar energy scaling of the ionization normalized yield along the major polarization axis and for the creation of a characteristic low-energy structure in photoelectron spectrum.

At large ellipticities, the main CF effect is due to initial Coulomb disturbance at the exit of the ionization tunnel. The initial Coulomb disturbance, as our estimates in Eqs. (\ref{w_i1}) and (\ref{w_i0}) show, enhances the ionization yield. This is because the electrons are tunneled out at larger laser fields or with smaller initial transverse momentum when the initial CF is taken into account for the electron  drifting along  m.p.a. The enhancement factor is shown to be sharply pronounced at intermediate ellipticities when both of the above mentioned enhancement mechanisms contribute. In this region of ellipticity,  the yield is enhanced by an order of magnitude due to CF. 
\\

\section*{ACKNOWLEDGMENT} 

The authors acknowledge valuable discussions with C. H. Keitel.

\newcommand{\noopsort}[1]{} \newcommand{\printfirst}[2]{#1}
\newcommand{\singleletter}[1]{#1} \newcommand{\switchargs}[2]{#2#1}


\begin{thebibliography}{99}

\bibitem{Schafer_1993} K. J. Schafer, B. Yang, L. F. DiMauro, K. C. Kulander, Phys. Rev. Lett. {\bf 70}, 1599. (1993)
\bibitem{CorkumPRL93}  P. B. Corkum, Phys. Rev. Lett.   {\bf 71}, 1994 (1993).

\bibitem{Becker_review} W. Becker \textit{et al.}, Adv. Atom. Mol. Opt. Phys. \textbf{48}, 36 (2000).

\bibitem{Agostini_review} P. Agostini and L. F. DiMauro, Rep. Prog. Phys. \textbf{67}, 813 (2004).

\bibitem{Walker} B. Walker, B. Sheehy, L. F. DiMauro, P. Agostini, K. J. Schafer, and K. C. Kulander, Phys. Rev. Lett. \textbf{73}, 1227 (1994).

\bibitem{Moshammer} R. Moshammer, B. Feuerstein, W. Schmitt, A. Dorn, C. D. Schr\"oter, J. Ullrich, H. Rottke, C. Trump, M. Wittmann, G. Korn, K. Hoffmann, and W. Sandner, Phys. Rev. Lett. \textbf{84}, 447 (2000).

\bibitem{Feuerstein} B. Feuerstein, R. Moshammer, and J. Ullrich, J. Phys. B \textbf{33}, L823 (2000).

\bibitem{DiChiara} A. D. DiChiara, E. Sistrunk, C. I. Blaga, U. B. Szafruga, P. Agostini, and L. F. DiMauro, Phys. Rev. Lett. \textbf{108}, 033002 (2012).

\bibitem{Shvetsov} N. I. Shvetsov-Shilovski, S. P. Goreslavski, S. V. Popruzhenko, and W. Becker, Phys. Rev. A \textbf{77}, 063405 (2008).

\bibitem{Eberly} X. Wang and J. H. Eberly, Phys. Rev. Lett. {\bf 105}, 083001 (2010). 

\bibitem{Wang} X. Wang and J. H. Eberly, New J. Phys. {\bf 12}, 093047 (2010).

\bibitem{Uzer} F. Mauger, C. Chandre, and T. Uzer, Phys. Rev. Lett. {\bf 105}, 083002 (2010).

\bibitem{BrabecIvanov} T. Brabec, M. Yu. Ivanov, and P. B. Corkum, Phys. Rev. A \textbf{54}, R2551 (1996).

\bibitem{YudinIvanov} G. L. Yudin and M. Yu. Ivanov, Phys. Rev. A {\bf 63}, 033404 (2001).

\bibitem{Villeneuve} D. Comtois \emph{et al.},  J. Phys. B {\bf 38}, 1923 (2005).

\bibitem{Shvetsov-Shilovski} N. I. Shvetsov-Shilovski, S. P. Goreslavski, S. V. Popruzhenko and W. Becker, Laser Phys. \textbf{19}, 1550 (2009).

\bibitem{DiMauroNP09} C. I. Blaga \textit{et al.}, Nature Phys. {\bf 5}, 335 (2009).

\bibitem{Catoire} F. Catoire \emph{et al.}, Laser Phys. {\bf 19}, 1574 (2009).

\bibitem{XuPRL09} W. Quan \emph{et al.}, Phys. Rev. Lett. {\bf 103}, 093001 (2009).

\bibitem{Liu} C. Liu and K. Z. Hatsagortsyan, Phys. Rev. Lett. {\bf 105}, 113003 (2010).

\bibitem{Yan} T.-M. Yan, S. V. Popruzhenko, M. J. J. Vrakking, D. Bauer, Phys. Rev. Lett. {\bf 105}, 253002 (2010).

\bibitem{Liu_JPB} C. Liu and K. Z. Hatsagortsyan, J. Phys. B {\bf 44}, 095402 (2011).

\bibitem{Telnov} D. A. Telnov and  Shih-I Chu, Phys. Rev. A \textbf{83}, 063406 (2011).

\bibitem{Chu} Z. Zhou and  Shih-I Chu, Phys. Rev. A \textbf{83}, 033406 (2011).

\bibitem{Rost} A. K\"astner, U. Saalmann, J. M. Rost, Phys. Rev. Lett. {\bf 108}, 033201 (2012).

\bibitem{Burgdorfer} C. Lemell,  \textit{et al.}, arXiv:1109.0607v1 [physics.atom-ph] (2011).

\bibitem{Bucksbaum} P. H. Bucksbaum, M. Bashkansky, and T. J. McIlrath, Phys. Rev. Lett. {\bf 58}, 349 (1987).

\bibitem{Bashkansky} M. Bashkansky, P. H. Bucksbaum, and D. W. Schumacher, Phys. Rev. Lett. {\bf 60}, 2458 (1988).

\bibitem{Paulus_98} G. G. Paulus \textit{et al.}, Phys. Rev. Lett. {\bf 80}, 484 (1998).

\bibitem{Paulus_00} G. G. Paulus \textit{et al.}, Phys. Rev. Lett. {\bf 84}, 3791 (2000).

\bibitem{Becker_Paulus} W. Becker \textit{et al.}, Laser Phys.  {\bf 8}, 56 (1998).

\bibitem{Mittleman} P. Krsti\'c and M. H. Mittleman, Phys. Rev. A \textbf{44}, 5938 (1991).

\bibitem{Ehlotzky} A. Jaro\'n, J. Z. Kami\'nski and F. Ehlotzky, Opt. Commun. \textbf{163}, 115 (1999).

\bibitem{Starace} B. Borca, M. V. Frolov, N. L. Manakov, A. F. Starace, Phys. Rev. Lett. {\bf 87}, 133001 (2001).

\bibitem{Goreslavski}  S. P. Goreslavski, G. G. Paulus, S. V. Popruzhenko, N. I. Shvetsov-Shilovski, Phys. Rev. Lett. {\bf 93}, 233002 (2004).

\bibitem{Popruzhenko}  S. V. Popruzhenko, G. G. Paulus and D. Bauer, Phys. Rev. A \textbf{77}, 053409 (2008).

\bibitem{Popruzhenko1} S. V. Popruzhenko, and D. Bauer, J. Mod. Opt. \textbf{55}, 2573 (2008).

\bibitem{PPT}  A. M. Perelomov, V. S. Popov, and V. M. Terent'ev, Zh. Exp. Theor. Fiz. \textbf{52}, 514 (1967)
[Sov. Phys. JETP {\bf 25}, 336 (1967)]; 

\bibitem{ADK} M. V. Ammosov, N. B. Delone, and V. P. Krainov, Zh. Exp. Theor. Fiz.  {\bf 91}, 2008 (1986) [Sov. Phys. JETP  {\bf 64}, 1191 (1986)].

\bibitem{Landau77} L. D. Landau and E. M. Lifshitz, \emph{Quantum Mechanics} (Pergamon, Oxford, 1977) p. 293.

\bibitem{Landau_Mechanics}  L. D. Landau and E. M. Lifshitz, \emph{Mechanics} (Pergamon, Oxford, 1993) p. 170.

\bibitem{Usp} V. S. Popov, Usp. Fiz. Nauk  \textbf{174}, 921 (2004)  [Phys. Usp. \textbf{47}, 855 (2004)].

\bibitem{Mur} V. D. Mur, S. V. Popruzhenko, and V. S. Popov, JETP {\bf 92}, 777 (2001) [Zh. Exp. Theor. Fiz.\textbf{ 119}, 893 (2001)]; 


\end{thebibliography}
\end{document}